\documentclass[10pt]{article}

\usepackage{graphicx}
\usepackage{dcolumn}
\usepackage{bm}

\input{tcilatex}

\begin{document}

\title{Nuclear spintronics}
\author{Israel D. Vagner \\
\\
\small{\textit{RCQCE at  Department of Communication Engineering,}} \\
\small{\textit{Holon Academic Institute of Technology, 52 Golomb Str., Holon 58102, Israel}} \\
\small{\textit{and}} \\
\small{\textit{Grenoble High Magnetic Fields Laboratory, }}\\
\small{\textit{Max-Planck-Institute f\"{u}r Festk\"{o}rperforschung and CNRS, }}\\
\small{\textit{25 Avenue des Martyrs, BP166, F38042, Grenoble, Cedex 9, France }}}
\date{}
\maketitle

\begin{abstract}
The electron spin transport in condensed matter, \textbf{Spintronics}, is a
subject of rapidly growing interest both scientifically and from the point
of view of applications to modern and future electronics. In many cases the
electron spin transport cannot be described adequately without accounting
for the hyperfine interaction between electron and nuclear spins. Here, the
progress in physics and applications of these phenomena will be reviewed.
\end{abstract}
\bigskip

\section{Introduction}

Hyperfine interactions between the nuclear and electron spins play a crucial
role in a large variety of physical phenomena in normal and superconducting
metals \cite{AbragBk61,WintBk}, bulk semiconductors \cite{DP84} and in
quantum Hall and mesoscopic systems \cite{VM95}. Very recently the indirect
hyperfine interaction\ between nuclear spin qubits in semiconductor based
quantum computer proposals attracted growing attention \cite{PMV02}.

Application of high magnetic fields is a very powerful tool for studying the
electronic properties of a large variety of metals, semiconductors and
superconductors. Due to the Landau quantization of electron motion in
sufficiently strong magnetic field, most of the transport properties, such
as magnetization, conductivity etc. experience magnetic quantum oscillations
(QO) \cite{ShoenbBk}. The Landau quantization is most spectacularly manifest
in the electronic magneto-transport in low dimensional conductors. A
striking example are the celebrated quantum Hall effects \cite{KDP80}.

Apart from the anomalous enhancement of the well known QO in 2DES one
expects also strong QO in physical properties which are not sensitive to the
magnetic field in isotropic three dimensional metals. It was suggested in 
\cite{VME82} that in quasi-two-dimensional metals under strong magnetic
fields the nuclear spin lattice relaxation rate $T_{1}^{-1}$ should exhibit
strong magnetic oscillations .

This should be compared with the Korringa relaxation law \cite
{AbragBk61,WintBk} usually observed in three-dimensional normal metals,
which results in a magnetic field independent nuclear spin-relaxation rate.
This line of research seems to be useful in dense quasi-two-dimensional
electronic systems, as is the case, for example in synthetic metals (GIC's
etc.) and low-dimensional organic compounds.

A completely new line of research, the hyperfine interaction between nuclear
and electron spins in low dimensional semiconductors and nanostructures, has
been developed during the last decade both theoretically \cite{VM88}-\cite
{PVW03JPC} and experimentally \cite{DKSWP88}-\cite{Thurber02}.

Very recently the interplay between the nuclear spin ordering at ultra-low
temperatures and superconductivity have attracted rapidly increasing
theoretical \cite{DVW96}-\cite{KBB97} and experimental \cite{RHP97} -\cite
{OL97} interest.

Here I will outline the theoretical concepts and experimental achievements
in the new and quickly developing field of \textit{nuclear spintronics}.

\section{Quantized Nuclear Spin Relaxation Effect: \textit{QNSRE}.}

\subsection{Activation law for $1/T_{1}$ in QHE\ systems.}

In metals and doped semiconductors, usually, the leading contribution to the
spin - lattice relaxation process is due to the \textit{hyperfine Fermi
contact} interaction between the nuclear spins and the conduction electron
spins \cite{AbragBk61}. This interaction is represented by the Hamiltonian: $%
\hat{H}_{int}=-\gamma _{n}\hbar \vec{I}_{i}\cdot \vec{H}_{e}$ where $\gamma
_{n}$ is the nuclear gyromagnetic ratio, $I_{i}$ is the nuclear spin and $%
H_{e}$ is the magnetic field on the nuclear site, produced by electron
orbital and spin magnetic moments: $\vec{H}_{e}=-g\beta \sum_{e}{\frac{8\pi 
}{3}}\hat{s}_{e}\delta \left( \vec{r}_{e}-\vec{R}_{i}\right) $. Here $\vec{r}%
_{e}$ is the electron radius-vector, $\hat{s}_{e}$ is the electron spin
operator, $\beta ={\frac{e\hbar }{m_{0}c}}$ is the Bohr magneton and $g$ is
the electronic $g$-factor.

The nuclear spin-lattice relaxation rate $T{_{1}^{-1}}$, caused by the
hyperfine Fermi contact interaction between the nuclear spins and the
conduction electron spins, is related to the local spin-spin correlation
function through the equation:

\begin{eqnarray}
{T_{1}^{-1}}{} &=&{}{\frac{32\pi ^{2}}{9}}\gamma _{n}^{2}g^{2}\beta ^{2} 
\nonumber \\
&&\times {{\int }_{-{\infty }}^{{\infty }}e^{{-i{\omega }_{n}}{t}}}{\left\{ {%
<S}{^{+}\left( {\mathbf{R},t}\right) S^{-}}{\left( {\mathbf{R},0}\right) >}%
\right\} dt}
\end{eqnarray}

where S${^{+}\left( \mathbf{R}\right) }$, ${S^{-}\left( \mathbf{R}\right) }$
are the transverse components of the electron spin density operator at the
nuclear position \textbf{R}, and $\omega _{n}$ is the nuclear magnetic
resonance frequency.
%
%
\begin{figure}
\begin{center}
\includegraphics[width=2.4976in]{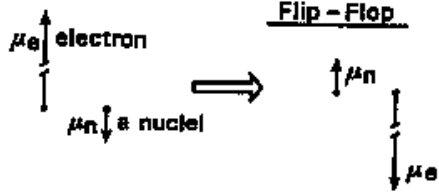}
\end{center}
\caption{Simultaneos electron and nuclear
spin flips induced by the contact hyperfine interaction.}
\end{figure}

The rate of the nuclear spin-relaxation in metals is, usually, proportional
to the temperature and to the square of the electronic density of states at
the Fermi energy (the Korringa law \cite{AbragBk61}). This follows from:

\begin{eqnarray}
{\frac{1}{T_{1}}} &\propto &\int_{0}^{\infty }\mid <i\mid V\mid f>\mid
^{2}\rho (E_{i})\rho (E_{f})f(E_{i})  \nonumber \\
&&\times \lbrack 1-f(E_{f})]\delta (E_{f}-E_{i}+\gamma _{N}H_{0}) 
\end{eqnarray}

At low temperatures: $f(E)[1-f(E)]\propto k_{B}T{\frac{\partial f}{\partial E%
}}$where $k_{B}T$ is the temperature in the energy units and we arrive at
the linear in temperature (Korringa) law

\begin{equation}
T_{1}^{-1}\sim k_{B}T\rho ^{2}\left( E_{F}\right)
\end{equation}
and $\rho \left( E_{F}\right) $ is the electron density of states at the
Fermi level.

In high magnetic fields and in systems with reduced dimensionality this
simple argumentation does not hold, since the electron spectrum acquires
field induced \cite{VM88} or size quantized \cite{VRWZ98,LGAA02} energy gaps.

It was conjectured in \cite{VM88} that in quantum Hall effect systems the
nuclear spin relaxation rate should have an activation behavior

\begin{equation}
T_{1}^{-1}\sim \exp \left\{ -\frac{\Delta \left( B\right) }{k_{B}T}\right\}
\label{T1exp}
\end{equation}
where $\Delta \left( B\right) $ is either $g\mu _{B}B$, the electron Zeeman
gap (odd filling factors) or $\hbar\omega _{c}$, the Landau levels gap (even
filling factors), instead of the usual Korringa law.

This unusual magnetic field dependence of the nuclear spin relaxation
reflects the fact that the energy gaps in the spectrum of two-dimensional
electrons in strong magnetic fields (either Zeeman splitting or the Landau
levels gap) are orders of magnitude larger than the nuclear Zeeman energy.
Indeed, the energy needed to reverse the spin of an electron in the external
magnetic field $H_{0}$ is $\Delta E_{el}=2g\mu _{B}H_{0}$, which is much
larger (by a factor of ${\frac{M_{n}}{m_{e}}\simeq }10^{3}$, $M_{n}$ and $%
m_{e}$ being the nuclear and free electron masses) than the energy $\gamma
_{n}H_{0}$ provided by reversing the nuclear spin. Therefore the delta
function in Eq. (2) can not be realized and the simultaneous spin flip of
the nuclear and the electron spins (flip-flop) Fig. 1, is severely
restricted by the energy conservation. The discreteness of the electron
spectrum will manifest, at finite temperatures, in an activation type of the
magnetic field dependence of the nuclear spin relaxation rate, $T_{1}^{-1}$,
Eq. (\ref{T1exp}), similar to that of the magnetoresistance $\rho _{xx}$ in
the QHE, \cite{VM88}.

In isotropic 3D electron systems, in a strong magnetic field : $\hbar \omega
_{c}>k_{B}T,$ where the kinetic energy of the electron motion perpendicular
to the field is quantized (the \textit{Landau levels}), the kinetic energy
of the electron motion parallel to the field should change in order to
ensure the energy conservation of the process. Thus, in the ''isotropic''
model, the electron spin - flip will be accompanied by a simultaneous change
of the Landau level and of the kinetic energy parallel to the field, E $_{z}$%
: $\Delta \epsilon _{z}=\hbar \omega _{c}(n^{\prime }-n)+$ $\ \gamma
_{n}H_{o}-\hbar \omega _{z}$. While this is impossible for an ideal 2D
system in a strong magnetic field, it may take place in
quasi-two-dimensional conductors, as is the case in superlattices, for
certain regions of parameters.

Because of the existence of energy gaps in the electron spectrum of a 2DES
under strong magnetic fields (the \textit{QHE} systems), finite nuclear spin
relaxation times $T_{1}$ could be expected only if 2DES is subjected to
different kinds of external potentials, such as short range impurities, \cite
{VM88,MV90,AMcD91}, long range potential fluctuations in a heterojunction 
\cite{IMV91,FMV91} and edge states \cite{VMS92}.

In sufficiently clean heterojunctions, however, where the FQHE and Wigner
crystallization could be observed, the mechanisms mentioned above are
extremely inefficient. At relatively high temperatures a phonon assisted
mechanism for relaxing the polarized nuclear spins can be operative \cite
{KVX94}.

Since the electron Zeeman energy gap reduces the effectiveness of the
contact interaction between the nuclear and electron spins, in very clean
limit the alternative relaxation channels, like the magnetic electron-nuclei
interaction (dipolar), may start to be operative \cite{OVD94}. The
dipole-dipole interaction does not conserve the total spin, and is not
sensitive, therefore, to the existence of the Zeeman gap in the electron
spectrum. In this process the spin angular momentum of nuclei is converted,
as a result of the interaction, to the orbital momentum of the electron gas.

\subsection{Spin-excitons}

Much of the recent attention paid to hyperfine interactions under conditions
of the quantum Hall effect is connected with correlation effects in 2DES.
This is based on the notion of a spin-exciton: the elementary excitation
over the Zeeman gap dressed by the Coulomb interaction \cite{BIE81}.This
results in a strong enhancement (up to a factor of 100, as is the case in
GaAs) of the effective $g(k)$ -factor.

Due to the Coulomb interaction the spin-excitons are bound states of
electron-hole pairs which, unlike the individual electrons or holes, can
propagate freely under the influence of a magnetic field due to their zero
electric charge. These elementary excitations are, therefore, chargeless
particles with a nearly parabolic dispersion in the low-k limit . At $k=0$
the gap is equal to the ''bare'' Zeeman splitting.

The energy spectrum of spin-excitons on the ground Landau level: n=0 is \cite
{BIE81}: $E_{ex}^{sp}(k)=\mid g\mid \mu _{B}H_{0}+\sqrt{\frac{\pi }{2}}[%
1-I_{0}({\frac{k^{2}}{4}})exp{{\frac{-k^{2}}{4}}}]{\frac{e^{2}}{\kappa a_{H}}%
.}$ In the parabolic approximation (small exciton momenta), the dispersion
relation reads: $E_{ex}^{sp}(k)\approx \mid g\mid \mu H_{0}+{\frac{k^{2}}{%
2m_{se}}}$where ${\frac{1}{2m_{se}}}\equiv {\frac{1}{4}}\sqrt{\frac{\pi }{2}}%
{\frac{e^{2}a_{H}}{\chi \hbar ^{2}}}$is the definition of the spin-exciton
mass.

The invariance of the energy gap with respect to the electron- electron
interaction is associated with the fact that in creating a
quasielectron-quasihole pair excitation at the same point in space (i.e.
with center of mass momentum $k=0$) the energy decrease due to the Coulomb
attraction is exactly cancelled by the increase in the exchange energy. Thus
the energy gap for the creation of a widely separated (i.e. with $%
k\rightarrow \infty $ ) quasielectron-quasihole pair (large spin-exciton) is
equal to the exchange energy associated with the hole.

\subsubsection{Long-range random potential}

Iordanskii et al. \cite{IMV91} have studied nuclear spin relaxation taking
into account the creation of spin-excitons in the flip-flop process. The
energy for the creation of a spin-exciton can be provided by the long range
impurity potential in a process, where the electron turns its spin while its
center of orbit is displaced to a region with lower potential energy. %
%

\begin{figure}
\begin{center}
\includegraphics[width=3.3347in]{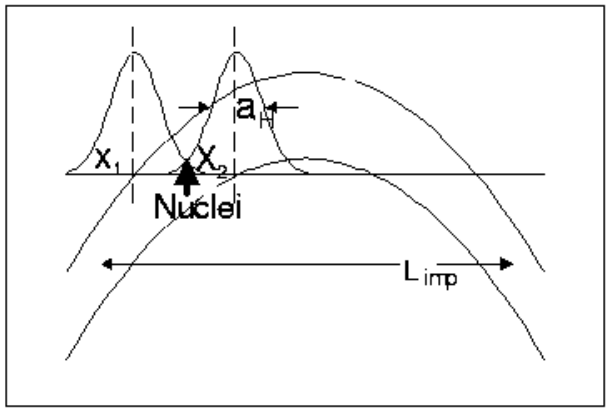}
\end{center}
\caption{Electron-nuclear spin flip flop
in the vicinity of the impurity potential.}
\end{figure}

As shown in Fig. 2, the overlap of the
initial and final location of the electron wave functions, centered at $%
x_{1} $ and $x_{2}$ respectively, is: $exp[-{\frac{(x_{1}-x_{0})^{2}}{a_{H}}}%
-{\frac{(x_{2}-x_{0})^{2}}{a_{H}^{2}}}]$. Here $x_{0}$ is the nuclear
position. Nuclear spin relaxation by the conduction electron spin in the
vicinity of a potential fluctuation is effective when the nuclear spin is
positioned in the region of the overlapping initial and final states of the
electron wave function.

The energy conservation in the spin-exciton creation process can be written
in the form: $\mu _{B}gH_{0}+E(p)=x\nabla U.$ This defines the gradient of
electric potential caused by the impurity, sufficient to create a
spin-exciton during a flip-flop process. The probability of finding such a
fluctuation is exponentially small: $exp[{\frac{-(\nabla U)^{2}}{2<\nabla
U^{2}>}}]$

The electronic DOS in high quality heterojunctions is defined by charged
impurities at a distance of the order of the spacer dimension. This
results,usually, in a gaussian random potential, which is smooth on the
scale of the magnetic length. Because the free spin-exciton energy is large
compared to the average potential value, it can be achieved only due to
rather large fluctuations of the random potential. It is possible,
therefore, to use the standard method of optimal fluctuation, to get the
expression for DOS \cite{IMV91}.

\subsubsection{Nuclear spin diffusion}

Apart from the direct nuclear spin relaxation, important information about
the electron system can be obtained from nuclear spin diffusion processes.
This is the case when the nuclear spins are polarized in a small part of a
sample as it was experimentally observed in \cite{KPW92,Wald94}.

To explain these experimental observations, Bychkov et al. \cite{BMV95b}
have suggested a new mechanism for indirect nuclear spin coupling via the
exchange of spin excitons. The spin diffusion rate from a given nuclear site 
$\vec{R}_{a}$ within the polarized region is proportional to the rate of
transition probability P($\vec{R}_{a}$) for the polarization of the nuclear
spin $\Downarrow $, located at $\vec{R}_{a}$ , to be transferred to a
nuclear spin $\Uparrow $, positioned at $\vec{R}_{b}$, outside the polarized
region, via the exchange of virtual spin excitons, Fig. 3. The
virtual character of the spin-excitons, transferring the nuclear spin
polarization, removes the problem of the energy conservation, typical for a
single flip-flop process. Furthermore, the virtual spin-excitons are neutral
entities, which can propagate freely in the presence of a magnetic field. In
this model the electron interactions play a crucial role: the kinetic energy
of a spin exciton is due to the Coulomb attraction between the electron and
the hole. Thus the proposed mechanism yields the possibility of transferring
nuclear spin polarization over a distance much longer than the magnetic
length $\ell _{B}$. The long range nature of this mechanism is of
considerable importance when the size of the region of excited nuclear
spins, $L_{ex}$, is much larger than the magnetic length $\ell _{B}$.

%
%

\begin{figure}
\begin{center}
\includegraphics[width=2.5495in]{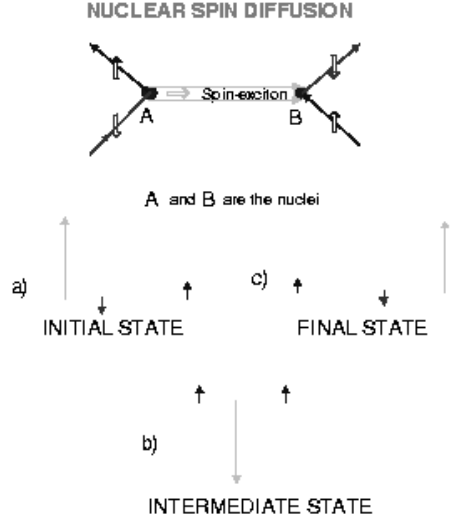}
\end{center}
\caption{Indirect interaction between two
nuclear spins via conduction electron spin.}
\end{figure}

As it is shown in \cite{BMV95b} , the potential of
the nuclear spin - spin interaction, mediated by the exchange of
spin-excitons, is a monotonic function of the distance between the two
nuclei with the asymptotics: $U\left( R_{ab}\right) \propto -\sqrt{\frac{d}{%
R_{ab}}}e^{-{\frac{R_{ab}}{d}}}$where $d\equiv {\frac{\ell _{B}}{2}}\sqrt{%
\frac{\epsilon _{c}}{\epsilon _{sp}}}$ . This is typical for the
interaction, mediated by the exchange of quasiparticles with an energy gap
at $q\rightarrow 0$ , as is the case for the spin-exciton dispersion. The
range $\Delta R$ of this potential is determined by the critical wave number 
$k_{0}={\frac{2}{\ell _{B}}}\sqrt{\frac{\epsilon _{sp}}{\epsilon _{c}}}$ as
it follows from the uncertainty principle: $\Delta R\cdot k_{0}\simeq 1$.
The negative sign of this interaction corresponds to attraction between the
nuclear spins and may cause, at sufficiently low temperatures, a
ferromagnetically ordered nuclear state in QHE systems.

\section{NONEQUILIBRIUM\ NUCLEAR\ SPIN\ POLARIZATION}

\subsection{Meso-Nucleo-Spinics}

Once the nuclear spins are completely polarized, they produce a hyperfine
field of the order of several Teslas, strongly influencing all the
electronic transport \cite{DKSWP88}-\cite{GJ96}. This may result in a series
of new nonequilibrium mesoscopic phenomena (\textbf{meso-nucleo-spinics}).
Among them are a) \textit{HABE-}the hyperfine field induced Aharonov- Bohm
effect \cite{VRWZ98}, b) \textit{HAHE- }the hyperfine field induced
Anomalous Hall effect \cite{Bedn98}, c) the nuclear spin polarization
induced quantum dots (\textit{NSPI QD)} \cite{FIPV02}and nanowires (\textit{%
NSPI NW) }\cite{PSVW02}.

The nonequilibrium nuclear spin polarization may strongly influence the
correlated electron states, resulting in creation of \textit{SkyrNuons }i.e.%
\textit{\ }skyrmions, localized by the nonuniform hyperfine field of
polarized nuclei Fig. 3.

\subsubsection{Hyperfine Aharonov-Bohm Effect}

The physics of the \textit{HABE}: the Aharonov-Bohm effect driven by the
hyperfine field of polarized nuclei, developed in \cite{VRWZ98}, can be
understood along the following lines.

Persistent currents (PC) in mesoscopic rings reflect the broken clockwise -
anticlockwise symmetry of charge carriers momenta caused, usually, by an
external vector potential. Experimentally PC are observed when an
adiabatically slow time dependent external magnetic field is applied along
the ring axis. The magnetic field variation results in an oscillatory
behavior of the diamagnetic moment, with the magnetic flux quantum $\Phi
_{0}=\frac{hc}{e}$, which is a manifestation of the Aharonov-Bohm effect (%
\emph{ABE}).

In a quantum ring with a nonequilibrium nuclear spin population the \textit{%
ABE}-like oscillations of PC with time may exist, during the time interval
of the order of nuclear spin relaxation time $T_{1}$, \cite{VRWZ98}, since
the hyperfine field breaks the spin symmetry of charged carriers. Combined
with a strong spin - orbital interaction (\textit{SOI}), in systems without
center of inversion \cite{BR84} this results in the breaking of the
rotational symmetry of diamagnetic currents in a ring. Under the
topologically nontrivial spatial nuclear spin distribution the hyperfine
field produces an adiabatically slow time variation of the Berry phase of
the electron wave function The time variation of this topological phase may
result in observable oscillations of a diamagnetic moment (the PC).

Once the nuclear spins are polarized, i.e. if $\left\langle \sum_{i}\mathbf{I%
}_{i}\right\rangle \neq 0$ , the charge carrier spins feel the effective
hyperfine field $B_{hypf}=B_{hypf}^{o}\exp \left( -t/T_{1}\right) $ which
lifts the spin degeneracy even in the absence of an external magnetic field.
In GaAs/AlGaAs one may achieve a spin splitting due to hyperfine field of
the order of one tenth of the Fermi energy \cite{Wald94}.

A typical term for a heterojunction \textit{SOI }is the Bychkov-Rashba term 
\cite{BR84}: $\widehat{H}_{so}=\frac{\alpha }{\hbar }\left[ \mathbf{\sigma }%
\times \mathbf{p}\right] \mathbf{\nu ,}$ where $\alpha =0.6\cdot 10^{-9}eVcm$
for holes, and $\alpha =0.25\cdot 10^{-9}eVcm$ for electrons, $\mathbf{%
\sigma }$ , $\mathbf{p}$ are the charge carrier spin and momentum and $%
\mathbf{\nu }$ is the normal to the surface. It can be rewritten in the form 
$\widehat{H}_{so}=\frac{e}{cm^{\ast }}\mathbf{pA}_{eff},$where $\mathbf{A}%
_{eff}=\frac{\alpha cm^{\ast }}{e\hbar }\left[ \mathbf{\nu }\times \mathbf{%
\sigma }\right] .$ Under the conditions of a topologically nontrivial
orientation of $\mathbf{A}_{eff}$ the wave function of a charge carrier
encircling the ring gains a phase shift similar to the one in an external
magnetic field as in the ordinary \textit{ABE}. This phase shift can be
estimated as follows: $2\pi \Theta =\frac{e}{\hbar c}\oint \mathbf{A}_{eff}d%
\mathbf{l}=\frac{m^{\ast }}{\hbar ^{2}}\left\langle \sigma (t)\alpha
\right\rangle \sim \frac{m^{\ast }\sigma (t)\alpha }{\hbar ^{2}}L,$where $L$
is the ring perimeter. To observe the oscillatory PC connected with the
adiabatically slow time-dependent $\left\langle \sigma \left( t\right)
\right\rangle ,$ $L$ is supposed to be less than the phase breaking length.
Taking the realistic values for $L\approx 3\mu m$ and $\left\langle \sigma
\right\rangle \approx 0.05\div 0.1$, the estimate: $2\pi \Theta \sim 5\div
10 $ shows the experimental feasibility of this effect.

There is a marked difference between the periodical time dependence of
standard \textit{ABE} oscillations, which are usually observed under the
condition of linear time variation of the applied magnetic field, and the
hyperfine field driven oscillations which die off due to the exponential
time dependence of the nuclear polarization.

It is worth pointing out that the PC depends not only on $B_{hypf},$ it also
has an oscillating dependence on \emph{SOI} coupling parameter $\alpha $ 
\cite{LGAA02}.

\subsubsection{Hyper-anomalous Hall effect}

The anomalous Hall effect, (\textit{AHE}), is caused by the spin-orbit
interaction, \textit{SOI}, combined with carrier magnetization. Due to 
\textit{SOI}, electrons with their spin polarization parallel to the
magnetization axis will be deflected at right angles to the directions of
the electric current and of the magnetization while electrons with
antiparallel spin polarization will be deflected in the opposite direction.
Thus, if the two spin populations are not equal there appears a net current
in the transverse direction. Until now, studies of the anomalous Hall effect
have been limited to the case where the carrier magnetization is induced
either by the external magnetic field or by the ordering of the magnetic
impurities. The magnetic field, however, produces a much larger normal Hall
effect which makes experimental studies quite difficult.

Bednarski et al. \cite{Bedn98} have suggested that the hyperfine field
induced anomalous Hall effect , \textit{HAHE}, can be obtained under
conditions of strong nuclear spin polarization even in the absence of an
external magnetic field. In this model the role of the hyperfine field is to
split the conduction band, thus creating the magnetization of the carriers,
and to introduce a long time scale dependence into electronic spin
polarization.

It is outlined in \cite{Bedn98}, that the magnetic field enters the results
due to non - commutativity of the $\mathbf{k}$ - vector components, while
the contact hyperfine field is a fictitious magnetic field, acting only on
electronic spins, and is not connected with any real vector potential. Thus
the presence of polarized nuclei can not be reduced to a trivial and formal
replacement of $\mathbf{B}$ by $\mathbf{B+B}_{HF}(t).$ For example, the
hyperfine field does not appear explicitly in the cyclotron frequency $%
\omega _{c}$ while it can influence $\omega _{c}$ via a spin dependent
effective mass.

\subsubsection{Skyrmions and Skyr-Nuons}

Skyrmions, in QHE systems, are the topologically nontrivial spin excitations
around filling factor $\nu =1$ \cite{lee90} which arise as a condensate of
interacting spin excitons \cite{BMV96PRB}. The Coulomb interaction acts to
enlarge the Skyrmion size while the Zeeman splitting tends to collapse
Skyrmions. The interplay between these factors determines the final
distribution of spins within a Skyrmion, and its characteristic length
scales. The resulting radius $R$ corresponds to the region where both these
energies are of the same value, and grows weakly to infinity as the g-factor
goes to zero \cite{BKMV96}, thus reflecting the importance of the long range
Coulomb repulsion associated with the Skyrmion charge in the zero g-factor
limit.

In NMR experiments on Skyrmions \cite
{Barrett94,Barrett98a,KKE99,Vitkalov00,Barrett01}, the nuclear spins are
strongly polarized. The sample inhomogeneity will result in a strong
inhomogeneity of the hyperfine field and therefore spatially varying
electron Zeeman splitting. This may result in a strong localization of
skyrmions \cite{NN99,Barrett01}, resulting in the combined topologically
nontrivial electron-nuclear spin excitation, the Skyr-Nuon, Fig. 4. 
%
%

\begin{figure}
\begin{center}
\includegraphics[width=3.0848in]{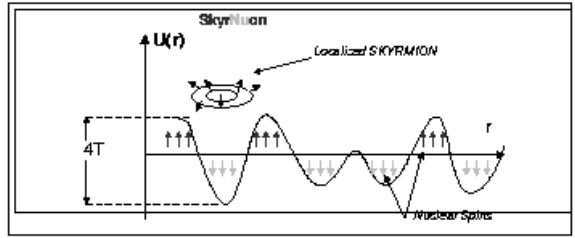}
\end{center}
\caption{Skyr-Nuon is a Skyrmion
localized by the interaction with the nuclear spins.}
\end{figure}

\subsection{Universal Residual Resistance}

The possibility that the hyperfine interaction between the conduction
electron spins and nuclear spins may result in hyperfine universal residual
resistivity \textit{(HURR) }in clean conductors at very low temperatures was
studied theoretically by Dyugaev et al. \cite{DVW96}. Apart from the
fundamental nature of this problem, the natural limitations on the mean free
path are decisive in semiconductor based high speed electronic devices, like
heterojunctions and quantum wells. The space periodicity of nuclei plays no
specific role, as long as the nuclear spins are disordered and act as
magnetic impurities with the concentration $C_{n}\approx 1$ . This
scattering is not operative at extremely low temperatures, in the $\mu K$
region when the nuclear spins are ferro- or antiferromagnetically ordered.

The residual ''nuclear ''resistivity is due to the Fermi (contact) hyperfine
interaction between the nuclear and the conduction electron spins:

\begin{equation}
V_{en}=-\frac{8\pi }3\mu _e\mu _h\Psi _e^2(0)\equiv \mu _nH_e  \label{Ven1}
\end{equation}
here $\mu _e$ and $\mu _h$ are the operators of the electron and nuclear
magnetic moments, $\Psi _e^2(0)\propto $ $Z$ is the value of the conduction
electron wave function on the nuclei with the nuclear charge $Z$ and $H_e$
is the magnetic field induced on nuclei by the electrons.

The estimate of $V_{en}$ in atomic units: $\hbar =m_{e}=e=1$ is: $%
V_{en}\approx Z\alpha ^{2}\frac{m_{e}}{M_{n}}Ry$ where $m_{e},M_{n}$ are the
electron and the nucleon masses, respectively; $Ry=27$ eV and $\alpha =\frac{%
1}{137}$ is the fine structure constant.

In metals the effective electron-nuclear interaction constant is \ $%
g_{n}\equiv \frac{V_{ne}}{\epsilon _{F}}\approx 10^{-7}Z\frac{Ry}{\epsilon
_{F}}$ where the Fermi energy $\epsilon _{F}$ varies in a wide interval $%
0.01\div 1$ . $g_{n}$ is $10^{-6}$ for $Li$ and $10^{-3}$ for the rare earth
metals.

The total residual resistivity is therefore a sum of the impurity $\rho
_{o}(T\rightarrow 0)\sim C_{o}$ and the nuclear spin $\rho _{n}(T\rightarrow
0)\sim g_{n}^{2}$ contributions: $\rho _{o}^{+}(0^{+})\approx \rho
_{oo}(C_{o}+g_{n}^{2})$. The nuclear contribution to resistivity starts to
be operative when the impurity concentration is $C_{o}\sim g_{n}^{2}$.

In the limit of an ideally pure ($C_{o}=0$) metal the universal residual
resistivity $\rho _{URR}$ is, therefore $\rho _{URR}\geq \rho _{oo}g_{n}^{2}$
and the mean free path is limited by $\frac{10^{-8}}{g_{n}^{2}}$ cm. This
yields $10^{4}$ cm in light atom conductors, as $Li,$ for example and $%
10^{-2}$ cm for the rare earth metals. It is interesting to note that in
zero-nuclear spin materials the \textit{HURR }should be absent.

In materials with the nuclear magnetic moments $I\neq \frac{1}{2}$, even
without external magnetic field, their $2I+1$ degeneracy is partially lifted
by quadrupole effects (in the case of cubic crystal symmetry the quadrupole
splitting of the nuclear levels may happen due to the defects \cite
{AbragBk61,WintBk,Rowlend60})and dislocations \cite{Averbuch59}. The
hyperfine nuclear contribution to $\rho _{n}(T) $ in this case will have a
different temperature dependence since $\mu _{n}H$ should be replaced by the
characteristic quadrupole splitting of energy levels.

While the normal metals have a quite similar electronic structure, the
experimentally observed temperature dependence of the dephasing time $\tau
_{\varphi }$ is quite different. This was shown in \cite{Mohanty99,Esteve1},
where the value of \ $\tau _{\varphi }$ was defined by the magnetoresistance
measurements of long metallic wires $Cu,Au,Ag$ in a wide temperature
interval $10^{-2}<T<10^{^{o}}$K. In $Cu$ and $Au$ wires \cite{Mohanty99}$%
\tau _{\varphi }$ saturates at low temperatures which contradicts the
standard theory \cite{AAK82}. Strangely enough the Ag wires do not show
saturation \cite{Esteve1} at the lowest temperatures, in accordance with 
\cite{AAK82}.

It is conjectured in \cite{DVW00CM} that the influence of the quadrupole
nuclear spin splitting on the phase coherence time $\tau _{\varphi }$ can be
a clue to this puzzle. Indeed, the nuclear spins of \ both $Cu$ and $Au$
have a strong quadrupole moment ($s=3/2$) and may act as inelastic two-level
scatterers once their degeneracy is lifted by the static impurities \cite
{Rowlend60}, dislocations \cite{Averbuch59}and other imperfections. It is
known that (see \cite{KA00} and references therein) the nondegenerate two
level scatterers may introduce inelastic phase breaking scattering of
conduction electrons.

This may be not the case for $Ag$ nuclei since their spin is $s=\frac{1}{2}$%
. In this case the quadrupole splitting of nuclei spins by imperfections is
negligible. In the absence of magnetic Zeeman splitting therefore the
nuclear spins in $Ag$ samples will act just as a set of elastic scatterers,
and the temperature dependence of $\tau _{\varphi}$ should obey the standard
theory \cite{AAK82}.

\subsection{Nuclear spins and superconducting order}

The problem of coexistence of the superconducting and magnetic ordering, in
spite of its long history \cite{Ginzb57}, is still among the enigmas of
modern condensed matter physics. Most of the theoretical and experimental
efforts were devoted to studies of the coexistence of electron
ferromagnetism and superconductivity. The possibility of a reduction of $%
H_{c}(T)$ by the nuclear ferromagnetism was outlined by Dyugaev et al. in 
\cite{DVW96}. Later on it was theoretically considered in more details \cite
{DVW97,KBB97}.

In \cite{DVW96} it was suggested that apart from the influence of the
''electromagnetic'' part of the polarized nuclear spins on the
superconducting order, the hyperfine coupling between the nuclear spins and
conduction electron spins may play a crucial role on the coexistence between
superconducting state and nuclear ferromagnetism \cite{DVW97}. Moreover, it
was conjectured in \cite{DVW02CM} that the hyperfine part of the
nuclear-spin-electron interaction may result in the appearance of a
nonuniform superconducting order parameter, the so called
Fulde-Ferrel-Larkin-Ovchinnikov state (FFLO) \cite{FF64}.

The FFLO state was thought originally to take place in superconductors with
magnetically ordered magnetic impurities \cite{FF64}. The main difficulty,
however, in the observation of the FFLO caused by magnetic impurity ordering
is in the simultaneous action of the ''electromagnetic'' and ''exchange''
parts of the magnetic impurities on the superconducting order. In most of
the known superconductors the ''electromagnetic'' part destroys the
superconducting order before the ''exchange'' part modifies the BCS
condensate to a nonuniform FFLO state.

The situation may change drastically in the case of nuclear spin
ferromagnetic ordering. Indeed, the nuclear magnetic moment $\mu _{n}=\frac{%
\hbar e}{M_{i}c},$ is at least three orders of magnitude smaller than the
electron Bohr magneton $\mu _{e}=\frac{\hbar e}{m_{o}c},$ so that the
''electromagnetic'' part of the nuclear spin fields is quite low, compared
to that of the magnetic impurities. On the other hand the ''exchange'' part
is strongly dependent on the nuclear charge $Z$. As it was shown in \cite
{DVW02CM}, in some materials the interplay between these two contributions
can be in favor of the ''exchange'' part, thus providing the necessary
conditions for the appearance of the FFLO.

\section{NUCLEAR\ SPINS\ AS\ QUBITS\ }

\subsection{Quantum Hall Quantum Computation}

The hyperfine interactions are believed to play a central role in solid
state electronics based realizations of future quantum computation devices 
\cite{PMV02}.

Privman et al. in \cite{PVK98} have proposed a quantum computer realization
based on hyperfine interactions between the conduction electrons and nuclear
spins embedded in a two-dimensional electron system in the quantum-Hall
effect. For modifications and improvements of this model see a recent review 
\cite{PMV02}.

The general idea is as follows: consider a chain of spin-$1/2$ nuclei in an
effectively two-dimensional heterojunction or quantum well subjected to a
strong magnetic field. The typical separation should be comparable to the
magnetic length $\ell_{H}=\sqrt{hc/eH}$, where $H$ is the applied magnetic
field, perpendicular to the 2D layer. This length is of the order of $100\AA$%
. The control of individual nuclear spins is by electromagnetic-radiation
pulses in the nuclear magnetic resonance (NMR) frequency range.
Differentiation between individual nuclear spins can be achieved by a
combination of several methods. First, one can use different nuclei.
Secondly, they can be positioned in different crystalline environments. The
latter can be controlled by implanting atoms and complexes into the host
material. Use of a magnetic field gradient could be also contemplated, but
there are severe limitations on the field variation owing to the need to
maintain the QHE electronic state. There is no apparent limit on how many
different spins can be arranged in a chain. It may also be appropriate to
utilize small clusters of nuclear spins, rather than individual spins. These
can be made coherent by lowering the temperature to the order of several $%
\mu $K [43].

Maniv et al. \cite{MBVW01} have suggested a physical process of preparing a
coherent state in QH ferromagnets. Let us assume that at time $t=-t_{0}<0$
the filling factor was tuned to a fixed value $\nu =\nu _{0}\neq 1$ and then
kept constant until $t=0$ . If $t_{0}\gg T_{2}\left( \nu _{0}\right) $ then
at $t=0$ the nuclear spin system is in the ground state corresponding to the
2D electron system at $\nu =\nu _{0}$. Suppose that at time $t=0$ the
filling factor is quickly switched ( i.e. on a time scale much shorter than $%
T_{2}\left( \nu _{0}\right) $ ) back to $\nu =1$ so that the nuclear spin
system is suddenly trapped in its instantaneous configuration corresponding
to $\nu =\nu _{0}\neq 1$ . Thus the nuclear spins for a long time $t$ ( $\gg 
$ $T_{2}\left( \nu _{0}\right) $ ) will find themselves almost frozen in the
ground state corresponding to the 2D electron system at $\nu =\nu _{0}$,
since $T_{2}\left( \nu =1\right) \gg T_{2}\left( \nu _{0}\right) $.

\subsection{Decoherence and Dephasing (T$_{2}$).}

Maniv et al., \cite{MBVW01}, have considered the effect of vacuum quantum
fluctuations in the QH ferromagnetic state on the decoherence of nuclear
spins. It was shown there that the virtual excitations of spin excitons \cite
{BIE81}, which have a large energy gap (on the scale of the nuclear Zeeman
energy) above the ferromagnetic ground state energy, lead to fast incomplete
decoherence in the nuclear spin system. It is found that a system of many
nuclear spins, coupled to the electronic spins in the 2D electron gas
through the Fermi contact hyperfine interaction, partially loses its phase
coherence during the short (electronic) time $\hbar /\varepsilon _{sp}$, \
even under the ideal conditions of the QHE, where both $T_{1}$, and $T_{2}$
are infinitely long. The effect arises as a result of vacuum quantum
fluctuations associated with virtual excitations of spin waves (or spin
excitons ) by the nuclear spins. The incompleteness of the resulting
decoherence is due to the large energy gap of these excitations whereas the
extreme weakness of the hyperfine interaction with the 2D electron gas
guarantees that the loss of coherence of a single nuclear spin is extremely
small.

\section{EXPERIMENTS}

The measurement of the nuclear spin-relaxation in heterojunctions is a
challenging experimental problem, since the direct detection of the NMR
signals in solids requires usually $10^{17}-10^{20}$ nuclei. The number of
nuclear spins interacting with the two-dimensional electrons is however,
much smaller: $10^{12}-10^{15}$.

The first successful measurements of the magnetic field dependence of $%
T_{1}^{-1}$ under QHE conditions were performed in a series of elegant
experiments by the K. von Klitzing group, \cite{DKSWP88}. Combining ESR and
resistivity measurement techniques they have observed the shifting of the
ESR resonance frequency by the hyperfine field of nonequilibrium nuclear
spin population, which is the well known Overhauser shift \cite
{AbragBk61,WintBk}. In this experiment the 2D electron Zeeman splitting is
tuned to the pumping frequency. The angular momentum gained by a 2DEG
electron, excited to the upper Zeeman branch, is then transferred to the
nuclear spins, thus creating a nonequilibrium nuclear spin population.

These measurements show a close similarity between the magnetic field
dependence of the nuclear spin-relaxation rate and the magnetoresistance in
quantum Hall effect, as it was suggested theoretically in \cite{VM88}, thus
demonstrating clearly the importance of the coupling of nuclear spins to the
conduction electron spins in the nuclear relaxation processes in these
systems.

Various experimental techniques were used since and in what follows we will
describe shortly the main developments and achievements in experimental
studies of the hyperfine coupling between the nuclear spins and the
electrons in QHE, mesoscopic and superconducting systems.

Another way of measuring the nuclear spin relaxation and diffusion in a
heterojunction under strong magnetic field by transport techniques
(spin-diode) was demonstrated by Kane et al., \cite{KPW92} . They have
reported measurements performed on ``spin diodes`` : junctions between two
coplanar 2DEG's in which $\nu <$l on one side and $\nu >$ 1 on the other.
The Fermi level E$_{F}$ crosses between spin levels at the junction. In such
a device the 2DEG is highly conducting except in a narrow region (with a
width of the order of several hundred angstroms) where $\nu =1.$

Wald et al. \cite{Wald94} presented the experimental evidence for the
effects of nuclear spin diffusion and the electron-nuclear Zeeman
interaction on interedge state scattering. Polarization of nuclear spins by
dc current has proven to be a rich source for new, not always yet
understood, phenomena. This is the case, for example of anomalous spikes in
resistivity around certain fractional filling factors, observed by
Kronmuller et al. \cite{Kronm99} and in resistively detected NMR \ in QHE
regime reported by Desrat et al. \cite{Desrat02}. \ Influence of
nonequilibrium nuclear spin polarization on Hall conductivity and
magnetoresistance was observed and studied in detail by Gauss et al. \cite
{GJ96}

In 1994 \cite{Barrett94} Barrett et al. observed, for the first time, a
sharp NMR signal in multi-quantum wells, using the Lampel \cite{Lampel68}
technique of polarizing the nuclear spins by optical pumping of interband
transitions with near-infrared laser light \textit{(OPNMR)}. Polarization of
nuclei results in a significant enhancement of the NMR sensitivity, since
the resonance in a two-level system results in equalizing their population.
The difference in the population is, obviously, maximal when the spins are
completely polarized.

Detailed studies of the Knight shift data suggested \cite{Barrett94} that
the usually accepted picture of electron spins, aligned parallel to the
external field, should be modified to include the possibility of
topologically nontrivial nuclear spin orientations, the Skyrmions. Optical
polarization of nuclear spins was used also as a tool for reducing the
Zeeman splitting of 2D electrons by Kukushkin et al. \cite{KKE99}. This
resulted in a noticeable enhancement of the skyrmionic excitations. Similar
results are reported by Vitkalov et al. \cite{Vitkalov00}.

Very recently, a modern ultra-sensitive ''standard'' NMR spin-echo technique
was employed to study the physics of quantum Hall effect \cite{Melinte99} in
GaAs/AlGaAS multi-quantum well heterostructures. The spin polarisation of
2DES in the quantum limit was investigated and the experimental data support
the noninteracting Composite Fermion model in the vicinity of the filling
factor $\nu =\frac{1}{2}$ . \ Using the same technique the polarization of
2DES near $\nu =\frac{2}{3}$ was investigated \cite{Freitag01}. It was shown
there that a quantum phase transition from a partially polarized to a fully
polarized state can be driven by increasing the ratio between the Zeeman and
Coulomb energies.

An amazing phenomenon, following from the hyperfine coupling between the
electron and the nuclear spins, is the giant enhancement of the low
temperature heat capacity of GaAs quantum wells near the filling factor $\nu
=1$ , discovered in 1996 by Bayot et al. , \cite{BGMSS96}. \ As other
thermodynamic properties, it experiences quantum oscillations, following
from the oscillatory density of states $D\left( E_{F}\right) $ at the Fermi
level.

At about $T=25mK$ , and in clean samples, Bayot et al. \cite{BGMSS96} have
observed anomalous deviations from the free electron model, in the specific
heat value (up to four orders of magnitude) for the filling factor in the
range $0.5\leq \nu $ $\leq 1.5$ . Their explanation is that in this interval
of parameters, the electron system couples strongly to nuclear spin system
with a concentration of several orders of magnitude larger than the electron
one.

This raises a question about the origin of the strong coupling between the
electron and the nuclear spins in the interval $0.5\leq \nu $ $\leq 1.5$ .
The guess is the skyrmions, since they are predicted to appear just in the
same interval of the filling factor. Additional support for this mechanism
is in the results of \cite{Melinte99}, where the disappearance of the
nuclear spin contribution to the heat capacity was reported, as the ratio
between the Zeeman and Coulomb energies exceeds a certain critical value.
The Zeeman splitting of electrons was modified in these experiments by
tilting the magnetic field.

A new very promising technique for measuring spatially varying nuclear spin
polarization within a GaAs sample is reported in \cite{Thurber02}. In the
force detected NMR (FDNMR) the sample is mounted on a microcantilever in an
applied magnetic field. A nearby magnetic particle creates a gradient of
magnetic field which exerts a force on the magnetized sample and triggers
the cantilever oscillations. FDNMR is capable to perform the magnetic
resonance imaging of the sample with a very high accuracy. \ This method can
be useful in defining the spatial distribution of nuclear spin polarization
in non homogeneous samples. This information may be crucial for
understanding different peculiarities of data obtained by previously
described methods.

A new world of the low-temperature physics of the hyperfine interactions in
superconducting metals opens in the $\mu K$ region, where the nuclear spins
start ordering \cite{RHP97}-\cite{OL97}. Reduction of the critical magnetic
field of superconductors by the ferromagnetic ordering of the nuclear spins
has been recently discovered by the Pobell group \cite{RHP97}. They have
studied the magnetic critical field $H_{c}(T)$ of a metallic compound $\
AuIn_{2}$ where the superconductivity sets up at $T_{ce}=0.207K.$ \ They
have observed, in $AuIn_{2}$, the nuclear spin ferromagnetic ordering at $%
T_{cn}=35\mu K.$ It was observed in these experiments that the magnetic
critical field $H_{c0}=14.5$ G is lowered by almost a factor of two at $%
T<T_{cn}.$ 

\bigskip
We acknowledge stimulating discussions with Yu, Bychkov, A.
Dyugaev, T. Maniv, Yu. Pershin, V. Privman and P. Wyder. This work
was supported by European Grants: HPRI-CT-1999-40013, EU-RTN:
HPRN-CT-2000-00157, INTAS 2001 -- 0791 and by a grant from Israel
Science Foundation founded by the Academy of Sciences and
Humanities.


\begin{thebibliography}{99}
\bibitem{AbragBk61}  A. Abragam, The Principles of Nuclear Magnetism
(Clarendon, Oxford, 1961); C.P. Slichter, \textit{Principles of magnetic
resonance} (Springer Verlag, Berlin, 3rd ed, 1991).

\bibitem{WintBk}  J. Winter, \textit{Magnetic Resonance in Metals}
(Clarendon Press, Oxford,  1971).

\bibitem{DP84}  See for a review: M.I. Dyakonov and V.I. Perel, \textit{%
Theory of optical spin orientation of electrons and nuclei in semiconductors}
(In: \textit{Modern Problems in Condensed Matter Sciences}, Ed. by F. Meier
and B.P. Zakharchenya, vol.\textbf{8}, p.11 (North Holland,  Amsterdam,
1984).

\bibitem{VM95}  See for a review: I.D. Vagner and T. Maniv, \textit{%
Hyperfine Interaction in Quantum Hall Effect Systems}, Physica B \textbf{204}%
, 141 (1995).

\bibitem{PMV02}  See for a review: V. Privman, D. Mozyrsky, and I.D. Vagner, 
\textit{Quantum computing with spin qubits in semiconductor structures,}
Comp. Phys. Comm. \textbf{146}, 331 (2002).

\bibitem{ShoenbBk}  D. Shoenberg, \textit{Magnetic Oscillations in Metals}
(Cambridge Univ. Press, Cambridge, 1984).

\bibitem{KDP80}  K. von Klitzing, G. Dorda, and M. Pepper, Phys. Rev. Lett. 
\textbf{45}, 494 (1980); D.C. Tsui, H.L. St\"{o}rmer, and A.C. Gossard,
Phys. Rev. Lett. \textbf{48}, 1559 (1982); R.B. Laughlin, Phys. Rev. Lett. 
\textbf{50}, 1395 (1983).

\bibitem{VME82}  I.D. Vagner, T. Maniv and E. Ehrenfreund, Solid State Comm.
(1982).

\bibitem{VM88}  I.D. Vagner and T. Maniv, Phys.Rev.Lett. \textbf{61}, 1400
(1988).

\bibitem{MV90}  T. Maniv and I.D. Vagner, Surf. Sci. \textbf{229}, 134
(1990).

\bibitem{AMcD91}  D. Antoniou and A. MacDonald, Phys. Rev. B \textbf{43},
11686 (1991).

\bibitem{IMV91}  S.V. Iordanskii, S.V. Meshkov, and I.D. Vagner, Phys. Rev.
B \textbf{44}, 6554 (1991).

\bibitem{FMV91}  V.I. Fal'ko, S.V. Meshkov, and I.D. Vagner, J. Phys.:
Condens. Matter \textbf{3}, 5079 (1991).

\bibitem{VMS92}  I.D. Vagner, T. Maniv, and T. Salditt, in:  \textit{High
Magnetic Fields in Semiconductor Physics III}, Ed. by G. Landwehr, p.131
(Springer Series in Solid State Sciences, \textbf{101}, Berlin-Heidelberg,
1992).

\bibitem{KVX94}  Ju.H. Kim, I.D. Vagner, and L. Xing, Phys. Rev. B \textbf{49%
}, 16777 (1994).

\bibitem{OVD94}  Yu.N. Ovchinnikov, I.D. Vagner, and A. Dyugaev, JETP Lett. 
\textbf{59}, 569 (1994).

\bibitem{BMV95b}  Yu.A. Bychkov, T. Maniv, and I.D. Vagner, Solid State
Comm. \textbf{94}, 61 (1995).

\bibitem{Bedn98}  H. Bednarski, V. Fleurov, and I.D. Vagner, Physica B, 
\textbf{256-258}, 641 (1998).

\bibitem{VRWZ98}  I.D. Vagner, A.S. Rozhavsky, P. Wyder, and A.Yu. Zyuzin,
Phys. Rev. Lett. \textbf{80,} 2417 (1998); V.A. Cherkasskiy, S.N.
Shevchenko, A.S. Rozhavsky, and I.D. Vagner, Low Temp. Phys. \textbf{25},
541 (1999).

\bibitem{PVK98}  V. Privman, I.D. Vagner, and G. Kventsel, Phys. Lett. A 
\textbf{236}, 141 (1998).

\bibitem{GV00}  V. Gurevich and I.D. Vagner, Physica B \textbf{284-288},
1876 (2000).

\bibitem{DVW00CM}  A.M. Dyugaev, I.D. Vagner, and P. Wyder, \textit{On the
electron scattering and dephasing by nuclear spins}, cond-mat/0005005 (2000).

\bibitem{SSI01}  I. Shlimak, V.I. Safarov, and I.D. Vagner, J. Phys.:
Condens. Matter, \textbf{13}, 6059 (2001).

\bibitem{MPG01}  D. Mozyrsky, V. Privman, and L. Glasser, Phys. Rev. Lett. 
\textbf{86}, 5112 (2001).

\bibitem{Girv01}  A. Mitra and S.M. Girvin, cond-mat/0110078 (2001).

\bibitem{MPV01}  D. Mozyrsky, V. Privman, and I.D. Vagner, Phys. Rev. B 
\textbf{63}, 085313 (2001).\qquad 

\bibitem{MBVW01}  T. Maniv, Yu. A Bychkov, I.D. Vagner, and P. Wyder, Phys.
Rev. B \textbf{64}, 193306(2001).

\bibitem{ApelBych01}  W. Apel and Yu.A. Bychkov, Phys. Rev. B \textbf{63},
224405 (2001).

\bibitem{Falko01}  S.I. Erlingsson, Y.V. Nazarov, and V.I. Fal'ko, Phys.
Rev. B \textbf{64}, 196306 (2001).

\bibitem{FIPV02}  V. Fleurov, V.A. Ivanov, F.M. Peeters, and I.D. Vagner,
Physica E \textbf{14}, 361 (2002).

\bibitem{PSVW02}  Yu.V. Pershin, S.N. Shevchenko, I.D. Vagner, and P. Wyder,
Phys. Rev. B \textbf{66}, 035303 (2002).

\bibitem{BRVW02}  E.V. Bezuglyi, A.S. Rozhavsky, I.D. Vagner, and P. Wyder,
Phys. Rev. B \textbf{66}, 052508 (2002).

\bibitem{LGAA02}  Y.B. Lyanda-Geller, I.L. Aleiner, and B.L. Altshuler,
Phys. Rev. Lett. \textbf{89}, 107602 (2002).

\bibitem{PVW03JPC}  Yu.V. Pershin, I.D. Vagner and P. Wyder,  J. Phys.:
Condens. Matter \textbf{15}, 997 (2003).

\bibitem{DKSWP88}  M. Dobers, K. von Klitzing, J. Schneider, G. Weimann and
K. Ploog, Phys. Rev. Lett., \textbf{61}, 1650 (1988); A.Berg, M. Dobers,R.R.
Gerhardts and K.v. Klitzing, Phys. Rev. Lett. \textbf{64}, 2563 (1990).

\bibitem{KPW92}  B.E. Kane, L.N. Pfeiffer and K.W. West, Phys. Rev. B 
\textbf{46}, 7264 (1992).

\bibitem{Wald94}  K. Wald, L.P. Kouwenhoven, P.L. McEuen, N.C. Van der
Vaart, and C.T. Foxon, Phys. Rev. Lett. \textbf{73}, 1011 (1994); D.C.
Dixon, K.R. Wald, P.L. McEuen and M.R. Melloch, Phys. Rev. B \textbf{56},
4743 (1997).

\bibitem{Barrett94}  S.E. Barrett, R. Tycko, L.N. Pfeifer and K.W. West,
Phys. Rev. Lett. \textbf{72}, 1368 (1994); S.E. Barret et al., Phys. Rev.
Lett. \textbf{75}, 4290 (1995); See for a review on ODNMR: R. Tycko et al.,
Science, \textbf{268}, 1460 (1995).

\bibitem{Barrett98a}  P. Khandelwal, N.N. Kuzma, S.E. Barrett, L.N.
Pfeiffer, and K.W. West, Phys. Rev. Lett. \textbf{81}, 673 (1998); A.E.
Dementyev, P. Khandelwal, N.N. Kuzma, S.E. Barrett, L.N. Pfeiffer, and K.W.
West, Phys. Rev. Lett. \textbf{83}, 5074 (1999); N.N. Kuzma, P. Khandelwal,
S.E. Barrett, L.N. Pfeiffer, and K.W. West, Science \textbf{281}, 5377, 686
(1999).

\bibitem{Barrett01}  P. Khandelwal, A.E. Dementyev, N.N. Kuzma, S.E.
Barrett, L.N. Pfeiffer, and K.W. West, Phys. Rev. Lett. \textbf{86}, 5353
(2001).

\bibitem{Kronm99}  S. Kronmuller, W. Dietsche, K. von Klitzing, G.
Denninger, W. Wegscheider, and M. Bichler, Phys. Rev. Lett. \textbf{82},
4070 (1999).

\bibitem{Desrat02}  W. Desrat, D.K. Maude, M. Potemski, J.C. Portal, Z.R.
Wasilevski, and G. Hill, Physica E \textbf{12}, 149 (2002).

\bibitem{GJ96}  N. Gauss, A.G.M. Jansen, M.H. Julien, Y. Fagot-Revurat, M.
Horvatic, and P. Wyder, Europhys. Lett., \textbf{49}, 75 (2000).

\bibitem{KKE99}  I.V. Kukushkin, K. v. Klitzing, and K. Eberl, Phys. Rev.
Lett. \textbf{60}, 2554 (1999).

\bibitem{Vitkalov00}  S. Vitkalov, C.R. Bowers, J. A. Simmons, and J.L.
Reno, Phys. Rev. B \textbf{61}, 5447 (2000).

\bibitem{Melinte99}  S. Melinte et al., Phys. Rev. Lett. \textbf{82}, 2764
(1999); S. Melinte et al., Phys. Rev. Lett. \textbf{84}, 354 (2000).

\bibitem{Freitag01}  N. Freitag, Y. Tokunaga, M. Horvatic, C. Berthier, M.
Shayegan, and L.P. Levy, Phys. Rev. Lett. \textbf{87}, 136801 (2001).

\bibitem{BGMSS96}  V. Bayot, E. Grivei, S. Melinte, M.B. Santos and M.
Shayegan, Phys. Rev. Lett. \textbf{76}, 4584 (1996).

\bibitem{Smet02}  J.N. Smet, R.A. Deutschmann, F. Ertl, W. Wegscheider, G.
Abstreiter, and K. von Klitzing, Nature \textbf{415}, 281 (2002).

\bibitem{Thurber02}  K.R. Thurber, L.E. Harrel, R. Feinchtein, and D.D.
Smith, Appl. Phys. Lett. \textbf{80}, 1794 (2002).

\bibitem{DVW96}  A.M. Dyugaev, I.D. Vagner, and P. Wyder, JETP Lett. \textbf{%
64}, 207 (1996).

\bibitem{DVW97}  A.M. Dyugaev, I.D. Vagner, and P. Wyder, JETP Lett. \textbf{%
65}, 810 (1997); \textit{ibid}, \textbf{73}, 411 (2001).

\bibitem{KBB97}  M.L. Kulic, A.I. Buzdin, and L.N. Bulaevskii, Phys. Rev. B 
\textbf{56}, R11415 (1997); E.B. Sonin, J. Low Temp. Phys. \textbf{110}, 411
(1998).

\bibitem{RHP97}  S. Rehmann, T. Herrmannsd\"{o}rfer, and F. Pobell, Phys.
Rev. Lett. \textbf{78}, 1122 (1997); M. Seibold, T. Herrmannsd\"{o}rfer, and
F. Pobell, J. Low Temp. Phys. \textbf{110}, 363 (1998); T. Herrmannsd\"{o}%
rfer, Physica B \textbf{280}, 368 (2000); T. Herrmannsd\"{o}rfer, S.
Rehmann, M. Seibold, and F. Pobell, J. Low Temp. Phys. \textbf{110}, 405
(1998); T. Herrmannsd\"{o}rfer and D. Tayurskii, J. Low Temp. Phys. \textbf{%
124}, 257 (2001).

\bibitem{RhKTLJRN01}  T.A. Knuuttila, J.T. Tuoriniemi, K. Lefmann, K.I.
Juntunen, F.B. Rasmussen, and K.K. Nummila, J. Low Temp. Phys. \textbf{123},
65 (2001).

\bibitem{OL97}  A.S. Oja and O.V. Lounasmaa, Rev. Mod. Phys. \textbf{69}, 1
(1997).

\bibitem{BIE81}  Yu.A. Bychkov, S.V. Iordanskii, and G.M. Eliashberg, Pis'ma
Zh.Eksp.Teor.Fiz.\textbf{33}, 152 (1981); Sov. Phys-JETP Lett. \textbf{33},
143 (1981); C. Kallin and B.I. Halperin, Phys. Rev. B \textbf{30}, 5655
(1984).

\bibitem{BR84}  Yu.A. Bychkov and E.I. Rashba, J. Phys. C: Solid State Phys. 
\textbf{17}, 6039 (1984).

\bibitem{lee90}  D.H. Lee and C.L. Kane, Phys. Rev Lett. \textbf{64}, 1313
(1990); S.L. Sondhi, A. Karlhede, S.A. Kivelson, and E.H. Rezayi, Phys. Rev.
B \textbf{47}, 16419 (1993); L. Brey, H.A. Fertig, R. Cote, and A.H.
MacDonald, Phys. Rev. Lett. \textbf{75}, 2562 (1995).

\bibitem{BMV96PRB}  Yu. A. Bychkov, T. Maniv, and I.D. Vagner, Phys. Rev. B 
\textbf{53}, 10148 (1996); See for a review: I.D. Vagner, Yu. A. Bychkov,
A.M. Dyugaev, and T. Maniv, \textit{Hyperfine interactions and spin textures
in quantum Hall systems}, Physica Scripta, T\textbf{66}, 158 (1996).

\bibitem{BKMV96}  Yu. A. Bychkov, A. Kolesnikov, T. Maniv, and I.D. Vagner,
J. Phys.: Condens. Matter \textbf{10}, 2029 (1998).

\bibitem{NN99}  A.J. Nederveen and Yu. V. Nazarov, Phys. Rev. Lett. \textbf{%
82}, 406 (1999).

\bibitem{Rowlend60}  T. Rowlend, Phys. Rev. \textbf{119}, 900 (1960).

\bibitem{Averbuch59}  P. Averbuch, F de Bergevin, and W. Mulers-Waryanth
CRAS, \textbf{249}, 2315 (1959); P. Averbuch, CRAS \textbf{253}, 2674 (1961).

\bibitem{Mohanty99}  P. Mohanty, Ann. Phys. (Leipzig) \textbf{8}, 549
(1999); M.E. Gershenson, Ann. Phys. (Leipzig) \textbf{8}, 559 (1999).

\bibitem{Esteve1}  A.B. Gougam, F. Pierre, H. Pothier, D. Esteve and N.O.
Birge, J. Low Temp. Phys.\textbf{\ 118}, 447 (2000).

\bibitem{AAK82}  B. Altshuler, A. Aronov and D. Khmelnitskii, J. Phys. C 
\textbf{15}, 7367 (1982); B. Altshuler and A.G. Aronov, in \textit{%
Electron-Electron Interaction in Disordered Systems}, edited by A.L. Efros
and M. Pollak (North-Holland, Amsterdam, 1985).

\bibitem{KA00}  V.E. Kravtsov and B.L. Altshuler, Phys. Rev. Lett. \textbf{84%
}, 3394 (2000); A. Zawadowski, Jan von Delft, and D.C. Ralph, Phys. Rev.
Lett. \textbf{83}, 2632 (1999).

\bibitem{Ginzb57}  V.L. Ginzburg, Sov. Phys. JETP \textbf{4}, 153 (1957).

\bibitem{DVW02CM}  A.M. Dyugaev, I.D. Vagner, and P. Wyder, \textit{Nuclear
ferromagnetism induced FFLO state}, cond-mat/0112286 (2001)

\bibitem{FF64}  P. Fulde and R.A. Ferrel, Phys. Rev. \textbf{135}, 1550
(1964); A.I. Larkin, Yu.N. Ovchinnikov, J. Exp. Theor. Phys., \textbf{47},
1136 (1964) (Sov. Phys.,- JETP, \textbf{20}, 762 (1965)).

\bibitem{Lampel68}  G. Lampel, Phys. Rev. Lett. \textbf{20}, 491 (1968); D.
Paget, G. Lampel, B. Sapoval, and V.I. Safarov, Phys. Rev. B \textbf{15},
5780 (1997).
\end{thebibliography}
\end{document}